\documentclass{article}[12pt]
\usepackage{everypage}
\usepackage{authblk}
\usepackage{graphicx}
\usepackage{etex}
\usepackage{float,verbatim}
\usepackage{color,ulem}
\usepackage{amssymb,amsmath}
\RequirePackage{xspace}
\def\be{\begin{equation}}
\def\ee{\end{equation}}
\def\bee{\begin{eqnarray}}
\def\eee{\end{eqnarray}}

\begin{document}
\title{ 
New analysis for the correlation between gravitational and neutrino detectors during  SN1987A}
{\small
\author{ P.Galeotti$^a$, G. Pizzella$^b$}
\affil{\small
${ }^a$Dipartimento di Fisica dell'Universit\`a di Torino e INFN Sez. di Torino\\
${ }^b$
 Istituto Nazionale di Fisica Nucleare,Laboratori Nazionali 
di  Frascati}
\maketitle


\abstract{
Two major problems, still associated with the SN1987A, are: a) the signals observed with the gravitational waves detectors, b) the duration of the collapse. Indeed, a) the sensitivity of the gravitational wave detectors seems to be small for detecting gravitational waves and, b) while some experimental data indicate a duration of order of hours, most theories assume that the collapse develops in a few seconds. 
Since recent data of the X-ray NuSTAR satellite show a clear evidence of an asymmetric collapse, we have revisited the experimental data recorded by  the underground and gravitational wave detectors running during the  SN1987A.
New evidence is shown that confirms previous results, namely that the data recorded by the gravitational wave detectors running in Rome and in Maryland are strongly correlated with the data of both the Mont Blanc and the Kamiokande detectors, and that the correlation extends over a long period of time (one or two hours) centered at the Mont Blanc time. This result indicates that also Kamiokande detected neutrinos at the Mont Blanc time, and these interactions were not identified because not grouped in a burst.
}

\section{Introduction}
SN 1987A was the only supernova visible at naked eye after the Kepler supernova (1604). At the time of this event, four underground detectors (Mont Blanc \cite{lsd}, Baksan \cite{baksan}, Kamiokande \cite{kamio}, IMB \cite{imb}) and two gravitational wave antennas (in Rome and in Maryland) were running.
The Mont Blanc Liquid Scintillation Detector (LSD in the following) was the only experiment designed to search for low energy neutrino interactions from stellar collapses (energy threshold $\sim$5 MeV). The Baksan Scintillation Telescope (BST in the following) was a multi-purpose cosmic ray detector with an energy threshold $\sim$10 MeV. The Kamioka Nucleon Decay Experiment (KND in the following) and the Irvine Michigan Brookhaven (IMB in the following) experiments (energy threshold $\sim$8 MeV and $\sim$25 MeV respectively) were designed to search for proton decay candidates by observing the Cerenkov light produced in the detector water by the particles originated in the decay. 
These four detectors were running at different depths underground, being the Mont Blanc LSD located deeper than the others (at the minimum depth of about 5,200 meters of water equivalent). This implies a much smaller background in LSD as compared with the other detectors, because of the much smaller flux of cosmic ray muons interacting in the rock around the LSD detector that produces a much smaller background of neutral particles entering inside the detector and imitating neutrino interactions.

One major problem associated with a supernova explosion is the duration of the inner core collapse. According to most theories of supernova explosion, the collapse develops in a few seconds but, as we will discuss here, all the experimental data of the collapse originating this supernova indicate a duration of order of hours. In addition, one should note that almost all theories do not take into account core rotation and magnetic fields, even if pulsars, i.e. a possible final result of the collapse, have the strongest magnetic field and the fastest rotation in the universe. However, some unconventional models based on fast rotation and fragmentation of the collapsing core have been suggested soon after the explosion to explain the experimental data from neutrino and gravitational waves detectors  \cite{derujula,stella,castagnoli,olgaim}.  But only the recent observations of the remnant of SN1987A made by NuSTAR (Nuclear Spectroscopic Telescope Array, a satellite launched by NASA on June 2012 to study the X-ray sky) show a clear evidence of an asymmetric collapse \cite{nustar}. These data, in particular the high resolution analysis of $^{44}$Ti lines, show a direct evidence of large-scale asymmetry in the explosion: \it the massive star exploded in a lopsided fashion, sending ejected material flying in one direction and the core of the star in the other. \rm The asymmetry of the explosion is an essential requirement in support of a collapse in two stages.

At the time of SN1987A, the four underground detectors recorded statistically significant signals but not at the same time, and the origin of this time difference is still not well understood, as we will discuss. The Mont Blanc LSD detector recorded on line, i.e. on real time at the occurrence, five interactions on February 23rd at 2h 52m UT, one day before the optical discovery of this supernova. The signal was so clear that as early as a few days later (February 28th) the IAU Circular n. 4323 reporting this detection was distributed to the astronomical community by the Central Bureau for Astronomical Telegrams of the International Astronomical Union. Several days later, after the raw data were analyzed and cosmic ray muons subtracted, KND reported detection of eleven interactions at 7h35m UT. IMB reported the detection of eight interactions at 7h35m UT, BST of five interactions at 7h36m UT and also LSD detected two interactions at 7h36m UT.

Because of the different energy thresholds and masses of these detectors it was clear very soon that there is no contradiction in the data: the first burst could be due to a large flux of low energy neutrinos and the second one to a small flux of high energy neutrinos. Indeed the five low energy pulses detected by the scintillator of LSD with visible energies in the range between 5.8 and 7.8 MeV, correspond to visible energies at the limit to be detected by KND and BST, and are not detectable by IMB.

Furthermore the gravitational wave detectors in Rome and Maryland (GWR and GWM respectively in the following) recorded several signals in time coincidence between them and with the event detected by LSD at 2h52m UT\footnote{We remark that the correlation was observed in two independent analysis of the data recorded by \bf three \rm different detectors running at intercontinental distances, namely: a first analysis (a) among LSD, Rome and Maryland and a second analysis (b) among Kamiokande, Rome and Maryland. }. This unusual activity preceded the LSD signals by 1.1-1.2 s, with an absolute systematic error in timing of the order of 0.5 s. This observation was not expected because the sensitivity of the interaction of gravitational waves with the detectors seemed to be too small for detecting gravitational waves presumably produced by this extragalactic supernova. Indeed the classical cross-section for the interaction of  gravitational waves  with matter is far below that needed to detect GW \cite{ruffini,weinberg,nc}. 
However, there are considerations based  on the idea that cooperative effects might occur in the detector pushing the cross-section to much larger values \cite{prepa,moleti}. We wish to recall that the resonant bar gravitational wave detectors are also sensitive to particles \cite{cosmici}, different from gravitational waves.

In this paper we have reexamined a particular aspect of the analysis employed in the above cited literature, finding   a strong enforcement of the correlation between neutrino and GW detectors. The scheme of the paper is the following:
\begin{enumerate}
\item{we describe the correlation method used in our past analysis}
\item{we comment on some criticism which was put forward on our method}
\item{we point out a particular aspect found in correlating the neutrino and GW detectors, namely that both of them show the greatest correlation on 23rd at 2h 52m, at the LSD  five neutrino time}
\item{ we improve this result by making the two correlations completely independent one from the other one}
\item{we go throughout a new four detector correlation analysis and find a confirmation of previous result}
\item{we estimate the statistical uncertainty of our result}
\end{enumerate}

We do not  describe other results obtained in the past years, namely
 the extraordinary correlation  \cite{sn4,texas,chu} found between LSD and KND and between LSD and BST during a period of order of 2 hours centered at 2h52m UT. These results  have been shown mainly in the papers \cite{lath1,sn1,sn2,sn4} and in the review paper \cite{rosen} where various correlations among all neutrino and GW detectors were described.

\section{The correlation algorithm}
This algorithm\footnote{Suggested by Sergio Frasca.}, called the \it net excitation method \rm \cite{schutz}  and described in detail in \cite{sn2}, is based on the idea to make use of  \underline{all available data} \rm in underground detectors, and not only those considered to be produced by neutrino interactions. Here we shall give a brief description of this method to compare the data available from three independent detectors.

Let us consider first the gravitational wave detectors located in Rome and in Maryland and the neutrino detector LSD located deep underground in the Mont Blanc Laboratory. We have three independent files of data, that we call \it events \rm for simplicity, even if we are well aware that most of these \it events \rm are just background triggers. Since we sample the data of the GW antennas every second, in one hour we have 3600 \it events \rm for the Rome detector, and 3600 \it events \rm for the Maryland detector. These \it events \rm are all independent one from each other, since the integration times of the filtering procedures \cite{boni} are much smaller than one second. We have also a variable number of \it neutrino events\rm, of order of 50 per hour for the LSD detector, most of them being triggers due to background local radioactivity.

Obviously the data of these three files are completely independent. We consider the sum
$
E_{RM}(t)=E_{R}(t)+E_{M}(t)
$
where $E_R$ and $E_M$ are the measured energies (also called energy innovations, in kelvin units) of the \it events \rm obtained with the Rome (RO) and the Maryland (MA) detectors at the same time t, 3600 values $E_{RM}(t)$ per hour.

Then we compute the sum $E(t)=\sum_i E_{RM}(t_i)$
where $t_i$ is the time of the $i^{th}$ \it event \rm of the LSD neutrino detector. The summation is extended over a given time interval (say one hour) in which $N_{\nu}$ \it events \rm of the neutrino detector (most of them certainly due to background) are present. 

The background for this algorithm is obtained by calculating $E(t_1,t_2)=\sum_j (E_{R}(t_{1j})+E_{M}(t_{2j}))$ at 2$N_{\nu}$  times $t_{1j}$ and $t_{2j}$ chosen randomly within the time interval, which has a $\chi^2$ distribution with many degrees of freedom and approaches a gaussian distribution. In one hour we have many more than $3600\times 3600$ independent values of $E(t_1,t_2)$.

Since we do not know whether the events we consider are real signals or background, already in our first paper \cite{sn2} we stressed that \it when we talk of g.w. or of neutrinos, we refer to the events recorded by the corresponding detectors, without neither presuming nor excluding that a part or all of these events are actually due to physical g.w. or physical neutrinos \rm.

Our analysis consists in comparing the value E(t) with the very large number of background values determined by considering non coincident signals RO and MA, observed at times uncorrelated with the neutrino \it events\rm. In absence of any real signal we expect that E(t) be just one of the many $E(t_1,t_2)$ background values and, on average, we expect that half of the background values be larger than E(t) and half be smaller.

An application of this method has been discussed at length in \cite{sn2}. The fig.13 of ref.\cite{sn2} shows that the maximum correlation between the data of LSD and the GW detectors occurs when the GW data precede the LSD data by 1.1 s.

With regard to the KND data, the time measurements have an absolute error of $\pm 1$ minute. However the coincidence with the IMB data requires a time correction of +7.7 seconds on the KND time, in order to make the two bursts, the KND and the IMB ones, coincide in time( see ref. \cite{schramm}).
While waiting for the next supernova, we have studied again in more detail the available experimental data in two distinct cases: a) using the data  RO, MA and LSD, b) using the data RO, MA and KND.

\section{Discussion on the \it net excitation \rm method}
We must comment on a criticism raised by Dickson and Schutz \cite{schutz} about this method.
They applied the method to two simulated files of events, one file simulating the data of a neutrino detector and the other one simulating the data of the gravitational wave detectors. This is not our case, since we use three independent files, one for a neutrino detector and two for two independent GW detectors

For the GW file they consider $N=N_1+N_2$ events, $N_1$ for the Rome detector and $N_2$ for the Maryland detector. They state that for calculating the background one has $N$ independent combinations, instead we think there are $N_1\times N_2$ possible combinations.

In the Dickson and Schutz paper they state that we used different methods for the Kamiokande data and for the Mont Blanc data. It is evident that we used the same  \it net excitation \rm method.

They criticized that we find the correlation at a fractional time, 1.1 s delay. But it is known that the results of many measurement subjected to quantization can have error well below the quantum.

These remarks and others \cite{commento} were sent to PRD  on 30 June 1995 as a comment on the Dickson and Schutz paper, but PRD refused to publish it without a reasonable explanation.

\section{A new unexpected result}
Among the several correlation analyses done in the past (see the cited literature) one result \cite{percorsi} appeared very striking to us. We had applied the net excitation method to periods of one hour, stepped by 0.1 hour, correlating in a first analysis RO, MA, LSD and, in a second analysis, correlating RO, MA, KND.

 It was found  that the correlation curves show an extremely similar behavior:   both the RO, MA, LSD and RO, MA, KND curves have the greatest correlation  at the time of the LSD event.
This result was never discussed at length. 

However, one objection can be raised, that the two correlation curves are not completely independent, because some of the RO-MA data used for the correlation with LSD may have been also employed  for the correlation with KND, since some of the LSD events occur at the same time of some of the KND events. 

In order to obtain correlation curves completely independent  one from each other, in this new analysis we have searched for coincidences between LSD and KND and eliminated the RO and MA data occurring at these times, order of a few per hour.

For the correlations GW-neutrino detectors we  apply again the \it net excitation method, \rm having eliminated from the LSD and KND lists of events those in coincidence. Again we have used moving time periods of one-hour stepped by 0.1 hour. We obtain the result shown in fig.\ref{corresenza}, where we have considered a delay of 1.1 s between the neutrino and the GW signals, as we had done in all our previous analyses \cite{sn2,sn4}. 
\begin{figure}
\includegraphics[width=1.0\linewidth]{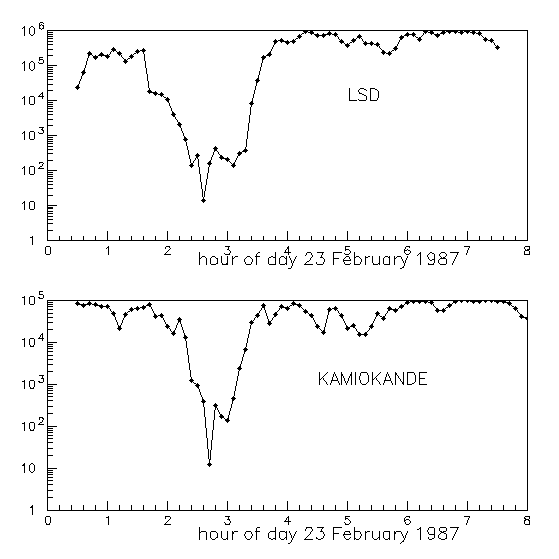}
 \caption{
The \it net excitation method \rm is applied on one-hour time periods moved in steps of 0.1 hour from 0 to 8 hours of 23 February, reported on the abscissa scale.  As in our previous analyses \cite{sn2,sn4} we have introduced a delay of 1.1 s between the neutrino and the GW signals.  On the ordinate scale we show the number of times, out of $10^6$ or $10^5$,  the GW background determinations are greater than the GW energy innovation obtained in correspondence of the neutrino events.  
        \label{corresenza} }
\end{figure}
The elimination of few events in common between LSD and KND has not changed substantially the behavior of the two correlation curves with GWR and GWM, as we find  by comparing these new results with those   \cite{percorsi} previously published.

We wish to stress that each one of the two curves has been obtained with independent  triple coincidences, the upper one RO-MA-LSD, the lower one RO-MA-KND. Similar correlations have been already found and discussed in our initial papers \cite{sn2,sn4}. 

In the present case, however, we have eliminated the coincidences  LSD-KND, thus the RO-MA data used for the upper curve are different from those used for the lower curve and we notice that the two correlations are still extremely similar.
This result again supports the idea that the supernova phenomenon lasted much more than a few seconds\footnote{That the phenomenon could have lasted much longer is discussed in  \cite{derujula,castagnoli,olgaim} }.

For calculating the  probability  that the result of fig.\ref{corresenza} was obtained by chance we consider the $i^{th}$ one-hour time periods. The  corresponding probabilities  that  the  correlation for LSD or KND occurred by chance are $\frac{n}{10^6}$ and $\frac{n}{10^5}$, where $n$ is the value on the ordinate axis of fig.\ref{corresenza}. Since the correlation GW-KND is due to data independent from the data for the correlation GW-LSD we can multiply the two   independent probabilities \cite{roe}
\be
p_i=\frac{n_{LSD}}{10^6}\times \frac{n_{KND}}{10^5}\times[1-log_e(\frac{n_{LSD}}{10^6}\times \frac{n_{KND}}{10^5})]
\label{totale}
\ee
In the case of random numbers we should have $p_i=0.5\times 0.5(1-log_e(0.5\times 0.5))=0.60$.
The neighbor one-hour time periods, stepped by 0.1 hour, are not independent one from each other. We plot the result in fig.\ref{proba} that gives a quantitative estimation of the probability that the two correlations  (RO-MA-LSD and RO-MA-KND) occurred at the same time.
\begin{figure}
\includegraphics[width=1.0\linewidth]{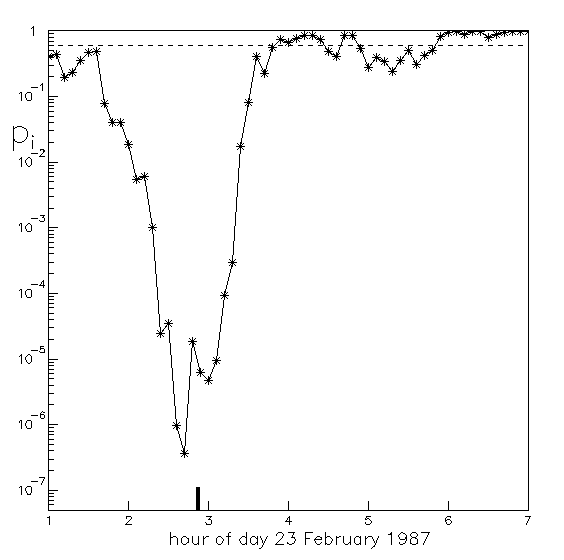}
 \caption{On the ordinate scale we show the probability $p_i$ that the two correlation curves (RO-MA-LSD and RO-MA-KND), calculated over periods of one-hour, stepped by 0.1 hour, give at each step similar probability values. The arrow indicates the time  $2^h52^m$ UT of the first \it real \rm neutrino interaction in the burst of five real events observed in the LSD experiment. The dashed line indicates the expected value in the case of absence of correlation.
         \label{proba} }
\end{figure}

This figure shows very clearly an extremely significant peak at the time of the Mont Blanc LSD burst of 5 interactions observed at 2h52m UT, with a probability  order of $10^{-6}$-$10^{-7}$ to be due to a random fluctuation.

We have considered periods of one hour as done in the past. In order to increase the time resolution we try to use periods of 30 minutes and the result  is shown in figures \ref{corresenza11} and \ref{probatre11}. We note that the best combined correlation occurs just at the time of the LSD burst of 5 interactions.
\begin{figure}
\includegraphics[width=1.0\linewidth]{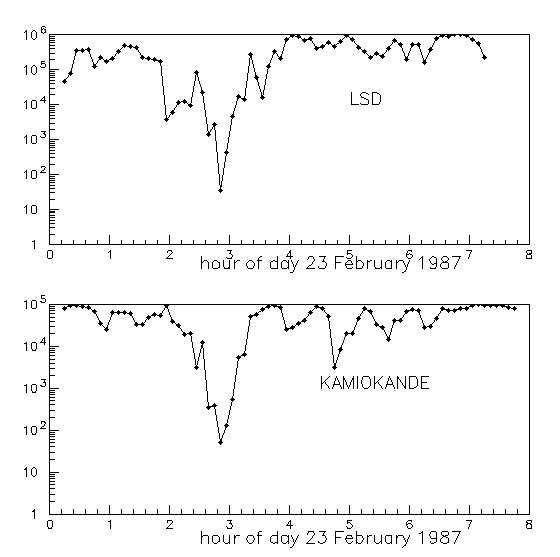}
 \caption{
The \it net excitation method \rm is applied on 30-minutes time periods moved in steps of 0.1 hour from 0 to 8 hours UT of 23rd February, shown on the abscissa scale. As in our previous analyses \cite{sn2,sn4} we have introduced a delay of 1.1 s between the neutrino and the GW signals. On the ordinate scale we show the number of times, out of $10^6$ or $10^5$ , the GW background determinations are greater than the GW energy innovation obtained in correspondence of the neutrino events.  
        \label{corresenza11} }
\end{figure}
\begin{figure}
\includegraphics[width=1.0\linewidth]{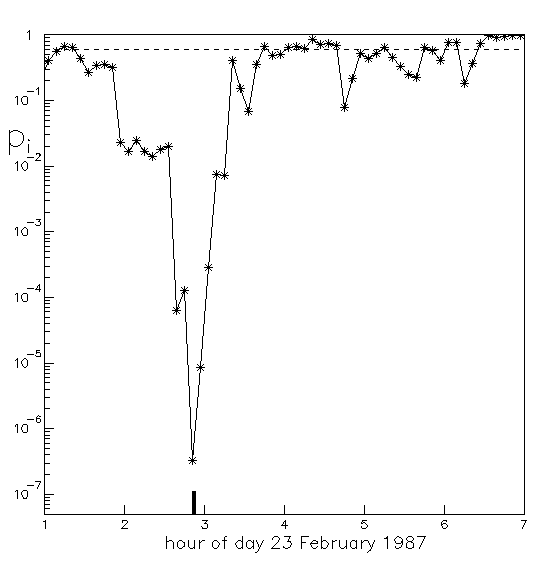}
 \caption{On the ordinate scale we show the probability $p_i$ that the two correlation curves (RO-MA-LSD and RO-MA-KND), calculated over periods of 30-minutes, stepped by 0.1 hour, give similar probability values at each step. The arrow indicates the time  $2^h52^m$ UT of the first \it real \rm neutrino interaction (out of five) observed by the LSD collaboration. The dashed line indicates the expected value in the case of absence of correlation.
        \label{probatre11} }
\end{figure}

The result, shown in figures \ref{corresenza11} and \ref{probatre11}, is very important to understand the physics of the collapse, because the best combined correlation occurs precisely at the time of the LSD burst of five neutrino interactions, i.e. 4.7 hours before the second burst.

\section{Quadruple correlation: LSD, KND, RO and MA}
We have applied the \it net excitation \rm method  also to search for correlations between GW detectors and a single list of events, obtained by adding together the LSD and KND data. Using one-half hour periods stepped by 0.1 hour we obtain the result shown in fig.\ref{quadru}
\begin{figure}
\includegraphics[width=1.0\linewidth]{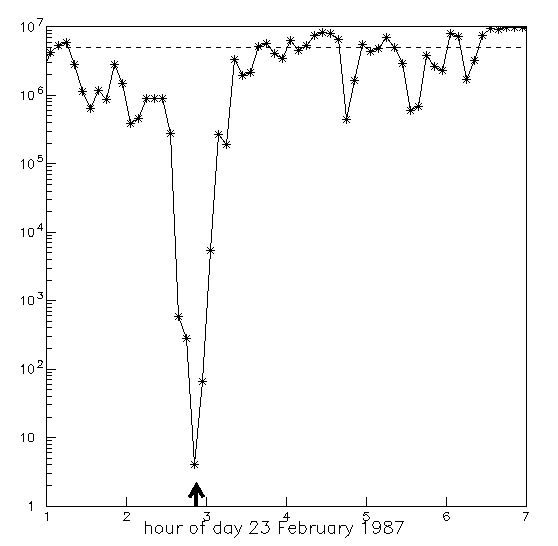}
 \caption{
The \it net excitation method \rm is applied on 30-minutes time periods moved in steps of 0.1 hour from 0 to 8 hours UT of 23rd February, shown on the abscissa scale. As in our previous analyses \cite{sn2,sn4} we have introduced a delay of 1.1 s between the neutrino and the GW signals. On the ordinate scale we show the number of times N, out of $10^7$ , the GW background determinations are greater or equal than the GW energy innovation obtained in correspondence of the neutrino \it  events \rm that includes both the LSD and the KND data. At the LSD time we have N = 4, corresponding to a probability of $4\times 10^{-7}$ that the correlation is accidental.The dashed line indicates the expected value in the case of absence of correlation.
        \label{quadru} }
\end{figure}

The similarity between fig.\ref{probatre11} and \ref{quadru} indicates the robustness of our probability estimation. During the period from 2h36m UT to 3h6m hour UT, that includes the LSD five-neutrino event at 2.87 hour UT, the sum of the 83 energy innovations of RO plus the 83 energy innovation of MA in coincidence with the 83 neutrino \it events\rm, divided by 83, was 74.349 kelvin, while the average background was 54.771 kelvin during the half an hour\footnote{The average noise of the RO detector is 28.6 K, that of MA is 22.1 K \cite{sn2}}. Fig.\ref{distri}  shows the distribution of ten million background determinations obtained during the above half an hour time with $\sigma$=4.08.  If the distribution of the background were gaussian we would have a difference between the signal and the average background equal to 74.349-51.771=5.5$ \sigma$ for a probability that our result was due to chance of $1.9\cdot 10^{-8}$. Our probability estimation of  $4\times 10^{-7}$ does not make any assumption about the experimental distribution.

The background determination were calculated by choosing randomly 83 energy innovations (32 from the LSD \it event  \rm times and 51 from the KND \it event \rm times) among the 1800 energy innovations of the RO detector and the 1800  energy innovation of the MA detector that are available in the half an hour period\footnote{The combinations of 83 objects among 1800 available from a GW detector are indeed a very large number. Equal number of combinations for the other GW detector}. We remark only three determinations of background, out of $10^7$, have value larger than  74.349 kelvin, that  one of the the energy innovations in coincidence with the 83 neutrino \it events\rm. This demonstrate the extraordinariness of the correlation.
\begin{figure}
\includegraphics[width=1.0\linewidth]{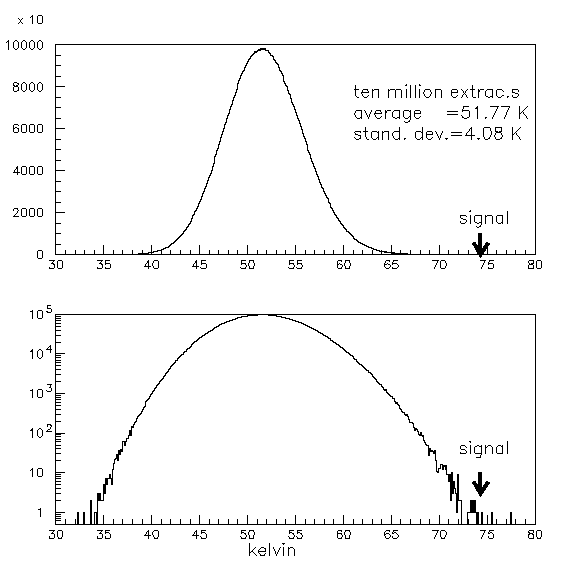}
 \caption{Background distribution.
The distribution in linear and logarithmic scale of ten million determinations of the background obtained by means of $10^7$ random extractions. The energy innovation in coincidence with the 83 neutrino \it events \rm (32 from LSD and 51 from KND) is 74,349 kelvin (the signal). From this graph we determine  the experimental probability to have the above signal by chance,  $4\times 10^{-7}$, about twenty times larger than in the case  the background distribution were gaussian.
        \label{distri} }
\end{figure}

\section{Discussion}
Since one of the comments we have often received in the past was that the LSD event is a statistical fluctuation in the data, \it a noise \rm, here again we stress that the Mont Blanc LSD signal is not only self consistent, but additionally supported by all the experiments running at that time, as shown in this paper and by the previous LSD KND and LSD BST correlation analyses \cite{sn4,chu}. 

We want to stress again that these new results have been obtained  by comparing six independent  files of events recorded in four different experimental apparatuses located at intercontinental distances one from the other,  furthermore exactly during the explosion of the only supernova visible at naked eye since 1604.

In more than thirty years of data taking from January 1985 in the Mont Blanc Laboratory with the LSD Experiment, and from June 1992 to nowadays in the Gran Sasso Underground Laboratory with the Large Volume Detector (LVD) with a mass of 1000 tons of liquid scintillator, no evidence of neutrino signals has been found \cite{lvd} a part the event printed on line in the night of February 23rd, 1987, just one day
before the detection of the only supernova visible at naked eye since 1604. This event
could certainly be a noise, but with very low  probability. We conclude that the chance that the results here discussed are due to a statistical fluctuation in the data is so small that a deep discussion is imposed, and we feel forced to try to find a possible explanation to this peculiar supernova that
was unusual under many other aspects \cite{hille} and not only for the recent NuSTAR
observations, even if this costs the use of a collapse scenario different from the presently scenarios.

Among the neutrino \it events \rm observed in the neutrino detectors very few have been attributed, by the researchers, to real neutrino interactions. However in our algorithm we make use of additional signals, other than those recognized as due to neutrino events. In our attempt to explain the experimental result described in this paper we suggest that many signals among those considered background (we recall a background with a counting rate per hour of about fifty for LSD and more than one hundred for KND) are real events due to neutrinos (or to any other exotic particles) that were not identified because  not grouped in a burst.

It is even more difficult to understand the cause of the signals in the GW detectors. As well known the predicted (but never measured) classical cross section of the bar-detectors for GW is at least one thousand times smaller than what would be needed to interpret the signals as due to GW, unless cooperative mechanisms are active in the bars, as already indicated in the Introduction \cite{prepa,moleti}, or we were so lucky to intercept beamed fluxes of gravitational waves. However we believe that we must not discard experimental results simply because we do not have a model, or even an explanation for them. Again we feel we must proceed in our attempt to explain what it is, in any case, an extraordinary experimental result. 

There are still two crucial points to be discussed. The first one is the time lag in the correlation between the GW and the neutrino detectors, and the second point is the astrophysical scenario of the supernova explosion. As far as the first point is concerned, we have found a time lag of 1.1 s  . The absolute time error for the GW apparatus in Maryland is of the order of 0.1 s, while the absolute time error for the GW apparatus in Rome has been discussed at length in paper \cite{sn2}. We can have a possible systematic error of 0.5 s at most, which would reduce the time lag to about 0.6 s. The UT time of LSD is measured by a clock with an absolute time error of 2 ms. The absolute time error for KND is 1 minute (more than enough for a proton decay experiment) so that, in our analysis, we have corrected the timing of the KND data by assuming that the two bursts at 7h 35m UT of  KND and IMB (whose clock precision is 5 ms) start at the same time as usually done \cite{schramm}.

In conclusion, we find a non-zero time gap between the GW and the underground detectors. To explain this gap it is necessary that the particles responsible for the signals in the GW detectors be emitted by the star 1.1 s (or 0.6 s) before the particles detected in the neutrino detectors, or that they are emitted simultaneously but the particles interacting in the neutrino detectors have a larger mass and arrive later. In the case of simultaneous emission, from the observed delay of $1.1 \pm0.5$ s between the first  and the second signals, we deduce that the neutrino mass
should be in the interval from 2.3 eV to 3.9 eV, not much different from the lower
bound of 2 eV quoted in \cite{mass} and the upper bound of 5.8 eV quoted in \cite{vissani}. In both cases we must imagine that the collapse mechanism be such that the two types of particles be emitted more or less simultaneously during a period of one or two hours.

Since the two neutrino bursts detected at 2h52m UT and 7h35m UT extend over a few seconds (7 s for LSD and 10 s for KND) at different times we are forced to consider the following scenario: \\
a) during several hours particles and perhaps gravitational waves  are emitted by  SN1987A,\\
b) the 7 s burst in LSD and the 10 s burst in KND are due to events which have been detected  because recorded in a short time interval,\\
c) the long lasting correlation with the GW detectors is due to massive neutrinos (or to exotic particles), not identified from the background because not grouped together but distributed over long times.

Even if it seems difficult to attribute the signals of the GW detectors to real gravitational waves or to particles, the recent observation of NuSTAR, \cite{nustar} mentioned in the introduction, shows a direct evidence of a large-scale asymmetry in the explosion: \it the massive star exploded in a lopsided fashion, sending ejected material flying in one direction and the core of the star in the other. \rm This experimental result not only is an important evidence of the fragmentation of the core, that involves a long duration of the collapse \cite{castagnoli,olgaim}, but it is also an essential requirement for the emission of gravitational waves, eventually along a narrow beam as it happens for optical or radio emission of pulsars. Finally, we have the experimental information that the collapse was asymmetric, a very important information to understand the mechanism of the collapsing inner core of the massive blue supergiants Sanduleak -69.202. 

\section{Conclusion}
We are aware that the experimental results described in this paper do not follow  most  theories on the supernova phenomenon, but our aim is not to
confirm theoretical models but to give experimental informations useful to explain this
peculiar supernova. 

 We wish to remark once again the great similarity between the correlations of the GW data with LSD and with KND, as shown in figs.\ref{corresenza} and \ref{corresenza11}, with a probability, quantified in  figs.\ref{proba} and \ref{probatre11}, lower than one in one million, to be occurred by chance just at the time of the LSD five-neutrino interaction. This result, reinforced by the analysis described in figs.\ref{quadru} and \ref{distri}, adds up to all previous correlation studies published in the above literature and induces us to think that they cannot be due to chance. 
 
 We conclude that this new analysis of the experimental data, obtained with the gravitational waves and neutrino detectors, strongly enforces the idea that between 2 and 4 hours UT of 23rd February 1987 both the neutrino and the GW bar-detectors were invested by intense fluxes of particles, not all of them  identified from the background because not grouped together but distributed over long times,  presumably originating from the SN1987A. As well known,  particles continued to arrive until 7.35 UT not detected by the gravitational waves detectors and with only two interactions detected by the LSD experiment al 7h 36 m UT. Obviously, something has changed in the structure and dynamics of the collapsing core in this 4.7 hours time interval between the first neutrino burst and the second one.
 
\section{Acknowledgment}
We dedicate this research-paper to Edoardo Amaldi, Carlo Castagnoli, Joseph Weber and George T. Zatsepin for their pursuing in forefront research and for their contribution to the realization of gravitational and neutrino detectors. We thank the Kamiokande, the LSD and the Rome Collaboration for having supplied to us their data.
We thank three anonymous referees for their criticism and suggestions  and Ugo Amaldi, Giovanni Vittorio Pallottino and Francesco Ronga for useful discussions.


\begin{thebibliography}{99}
\bibitem{lsd}M. Aglietta et al.
Europhys. Lett., \bf{3}\rm, 1315 (1987).
\bibitem{baksan}E. N. Alekseev et al Phys. Lett. \bf B205 \rm (1988) 209-214
\bibitem{kamio}K. S. HIirata et al.:  Phys.Rev.Lett.\bf{58}, \rm 1940 (1987)
\bibitem{imb}R. M. Bionta et al.: Phys. Rev. Lett., \bf 58, \rm1494 (1987).
\bibitem{derujula}A.DeRujula, Phys.Lett. \bf B193,\rm 514 (1987)
\bibitem{stella} L.Stella and A.Treves, Astronomy and Astrophysics, \bf 185,\rm L5-L6 (1987)
\bibitem{castagnoli} 
V.S. Berezinsky, C. Castagnoli, V. I. Dokuchaev , P Galeotti ,\\
 Nuovo Cimento \bf C11,\rm  287-303 (1988)
\bibitem{olgaim}V.S.Imshennik and O.G.Ryashskaya, Astronomy Letters, \bf{30}\rm ,14-31 (2004)
\bibitem{nustar} S.E. Boggs et al., Science, 348, 670, (2015)

\bibitem{ruffini}R.Ruffini and J.A.Wheeler
\it Relativistic cosmology and space platforms\rm \\
Proc. of ESRO Colloquium, Interlaken (1969)
\bibitem{weinberg} S.Weinberg "Gravitation and Cosmology"\\
John Wiley and Sons, Inc. New Kork (1972)
\bibitem{nc}G.Pizzella, Rivista del Nuovo Cimento \bf 5 \rm,369 (1975)

\bibitem{prepa}G.Preparata, Modern Physics Lett. A, 5 (1), 1-5, (1990)
\bibitem{moleti}R.Sisto and A.Moleti, Int. J. Mod. Phys. \bf D 13, \rm 625 (2004)

\bibitem{cosmici}
P. Astone  et al..   Astropart.Phys. \bf{30}, \rm 200 (2008) 

\bibitem{sn4}M.Aglietta et al.,  Nuovo Cimento \bf C14 \rm, 171-193 (1991)
\bibitem{texas} E.Amaldi et al.,  \it Texas Symposium on Relativistic Astrophysics\rm \\
 pages 561-576 (1989)
\bibitem{chu}A.E.Chudakov, \it Texas Symposium on Relativistic Astrophysics \rm \\
Volume  \bf 571 \rm 577-583, (1989)
\bibitem{sn2}M.Aglietta et al.,  Nuovo Cimento \bf C12\rm, 75,(1989) 
\bibitem{lath1}E.Amaldi et al., LA THUILE 
\it Results and Perspectives in Particle Physics,  \rm 59-68 (1987)
\bibitem{sn1}E.Amaldi et al., Europhys. Lett. \bf 3 \rm, 1325-1330 (1987)
\bibitem{rosen}G.Pizzella, \it Correlation among Gravitational Wave and Neutrino Detectors Data during SN1987A\rm\\
Jubilee Volume in Honour of Nathan Rosen. 
Editors: F.I. Cooperstock, L.P. Horowitz and J. Rosen. (1990)

\bibitem{schutz}C.A. Dickson and B.F. Schutz, Phys.Rev.\bf D51 \rm 2644-2668 (1995)

\bibitem{boni}P.Bonifazi et al. Nuovo: Cimento \bf C1,\rm 465 (1978).

\bibitem{schramm}D.N.Schramm, 
\it Neutrinos from Supernova SN 1987a \rm
Jun 1987 - 40 pages
Comments Nucl.Part.Phys. 17  239-278 (1987)\\
FERMILAB-PUB-87-091-A

\bibitem{commento}S.Frasca, G.V.Pallottino and G.Pizzella, \it Comment on; Reassessment of the reported correlation between gravitational waves and neutrinos associated with SN1987A, by C.D.Dickson and B.F.Schutz
PRD 51,2644 (1995) \rm )\\
Nota Interna 1088, 20 maggio 1997\\
Universit\`{a} deli Studi di Roma \it La Sapienza \rm and INFN.

\bibitem{percorsi} see for instance:  
G.Pizzella. \it Edoardo Amaldi and the Search for Gravitational Waves \rm
Italian Phys.Soc.Proc. \bf 100.\rm  31-58 (2010)\\
Conference: C08-10-23.2 Proceedings\\
G.Pizzella, \it Quaderni di Storia della Fisica, \rm 
n.7, pag 115-122, Societ\`{a} Italiana di Fisica (2000).


\bibitem{roe}B.P.Roe, \it Probability and Statistics in Experimental Physics, Springer \rm  pag.164 (2001)

\bibitem{hille}W. Hillebrandt, P. Hoflich, J.W.Truran and A. Weiss,\\ 
Nature, \bf {327}, \rm  597-600, 1987

\bibitem{mass}R. G. Hamish Robertson. Nuclear Physics B Proceedings Supplement 00 (2015) 1Ð6
\bibitem{vissani}Pagliaroli, Rossi-Torres and Vissani, Astroparticle Physics 33, 287-291, 2010

\bibitem{lvd} N.Y. Agafonova et al., The astrophysical Journal \bf 802,\rm 47 (2015)


\end{thebibliography}
\end{document}